\documentstyle[twoside,epsf]{article}

\catcode`\@=11
\long\def\@makefntext#1{
\protect\noindent \hbox to 3.2pt {\hskip-.9pt  
$^{{\eightrm\@thefnmark}}$\hfil}#1\hfill}		%CAN BE USED 

\def\@makefnmark{\hbox to 0pt{$^{\@thefnmark}$\hss}}	%ORIGINAL 
	
\def\ps@myheadings{\let\@mkboth\@gobbletwo
\def\@oddhead{\hbox{}
\rightmark\hfil\eightrm\thepage}   
\def\@oddfoot{}\def\@evenhead{\eightrm\thepage\hfil
\leftmark\hbox{}}\def\@evenfoot{}
\def\sectionmark##1{}\def\subsectionmark##1{}}

\oddsidemargin=\evensidemargin
\newcounter{sectionc}\newcounter{subsectionc}\newcounter{subsubsectionc}
\renewcommand{\section}[1] {\vspace{12pt}\addtocounter{sectionc}{1} 
\setcounter{subsectionc}{0}\setcounter{subsubsectionc}{0}\noindent 
	{\tenbf\thesectionc. #1}\par\vspace{5pt}}
\renewcommand{\subsection}[1] {\vspace{12pt}\addtocounter{subsectionc}{1} 
	\setcounter{subsubsectionc}{0}\noindent 
	{\bf\thesectionc.\thesubsectionc. {\kern1pt \bfit #1}}\par\vspace{5pt}}
\renewcommand{\subsubsection}[1] {\vspace{12pt}\addtocounter{subsubsectionc}{1}
	\noindent{\tenrm\thesectionc.\thesubsectionc.\thesubsubsectionc.
	{\kern1pt \tenit #1}}\par\vspace{5pt}}
\newcommand{\nonumsection}[1] {\vspace{12pt}\noindent{\tenbf #1}
	\par\vspace{5pt}}

\newcounter{appendixc}
\newcounter{subappendixc}[appendixc]
\newcounter{subsubappendixc}[subappendixc]
\renewcommand{\thesubappendixc}{\Alph{appendixc}.\arabic{subappendixc}}
\renewcommand{\thesubsubappendixc}
	{\Alph{appendixc}.\arabic{subappendixc}.\arabic{subsubappendixc}}

\renewcommand{\appendix}[1] {\vspace{12pt}
        \refstepcounter{appendixc}
        \setcounter{figure}{0}
        \setcounter{table}{0}
        \setcounter{lemma}{0}
        \setcounter{theorem}{0}
        \setcounter{corollary}{0}
        \setcounter{definition}{0}
        \setcounter{equation}{0}
        \renewcommand{\thefigure}{\Alph{appendixc}.\arabic{figure}}
        \renewcommand{\thetable}{\Alph{appendixc}.\arabic{table}}
        \renewcommand{\theappendixc}{\Alph{appendixc}}
        \renewcommand{\thelemma}{\Alph{appendixc}.\arabic{lemma}}
        \renewcommand{\thetheorem}{\Alph{appendixc}.\arabic{theorem}}
        \renewcommand{\thedefinition}{\Alph{appendixc}.\arabic{definition}}
        \renewcommand{\thecorollary}{\Alph{appendixc}.\arabic{corollary}}
        \renewcommand{\theequation}{\Alph{appendixc}.\arabic{equation}}
        \noindent{\tenbf Appendix \theappendixc #1}\par\vspace{5pt}}
\newcommand{\subappendix}[1] {\vspace{12pt}
        \refstepcounter{subappendixc}
        \noindent{\bf Appendix \thesubappendixc. {\kern1pt \bfit #1}}
	\par\vspace{5pt}}
\newcommand{\subsubappendix}[1] {\vspace{12pt}
        \refstepcounter{subsubappendixc}
        \noindent{\rm Appendix \thesubsubappendixc. {\kern1pt \tenit #1}}
	\par\vspace{5pt}}

\newcommand{\textlineskip}{\baselineskip=13pt}
\newcommand{\smalllineskip}{\baselineskip=10pt}

\def\eightcirc{
\begin{picture}(0,0)
\put(4.4,1.8){\circle{6.5}}
\end{picture}}
\def\eightcopyright{\eightcirc\kern2.7pt\hbox{\eightrm c}} 

\newcommand{\copyrightheading}[1]
	{\vspace*{-2.5cm}\smalllineskip{\flushleft
	{\footnotesize International Journal of Modern Physics B, #1}\\
	{\footnotesize $\eightcopyright$\, World Scientific Publishing
	 Company}\\
	 }}

\newcommand{\pub}[1]{{\begin{center}\footnotesize\smalllineskip 
	Received #1\\
	\end{center}
	}}

\def\abstracts#1#2#3{{
	\centering{\begin{minipage}{4.5in}\baselineskip=10pt\footnotesize
	\parindent=0pt #1\par 
	\parindent=15pt #2\par
	\parindent=15pt #3
	\end{minipage}}\par}} 

\renewenvironment{thebibliography}[1]			%ALL CHANGES DD 13/3/92
	{\frenchspacing
	 \ninerm\baselineskip=11pt
	 \begin{list}{\arabic{enumi}.}
	{\usecounter{enumi}\setlength{\parsep}{0pt}
	 \setlength{\leftmargin 12.7pt}{\rightmargin 0pt} %FOR 1--9 ITEMS
	 \setlength{\itemsep}{0pt} \settowidth
	{\labelwidth}{#1.}\sloppy}}{\end{list}}

\newcounter{itemlistc}
\newcounter{romanlistc}
\newcounter{alphlistc}
\newcounter{arabiclistc}

\newcommand{\fcaption}[1]{
        \refstepcounter{figure}
        \setbox\@tempboxa = \hbox{\footnotesize Fig.~\thefigure. #1}
        \ifdim \wd\@tempboxa > 5in
           {\begin{center}
        \parbox{5in}{\footnotesize\smalllineskip Fig.~\thefigure. #1}
            \end{center}}
        \else
             {\begin{center}
             {\footnotesize Fig.~\thefigure. #1}
              \end{center}}
        \fi}

\newcommand{\tcaption}[1]{
        \refstepcounter{table}
        \setbox\@tempboxa = \hbox{\footnotesize Table~\thetable. #1}
        \ifdim \wd\@tempboxa > 5in
           {\begin{center}
        \parbox{5in}{\footnotesize\smalllineskip Table~\thetable. #1}
            \end{center}}
        \else
             {\begin{center}
             {\footnotesize Table~\thetable. #1}
              \end{center}}
        \fi}

\def\@citex[#1]#2{\if@filesw\immediate\write\@auxout
	{\string\citation{#2}}\fi
\def\@citea{}\@cite{\@for\@citeb:=#2\do
	{\@citea\def\@citea{,}\@ifundefined
	{b@\@citeb}{{\bf ?}\@warning
	{Citation `\@citeb' on page \thepage \space undefined}}
	{\csname b@\@citeb\endcsname}}}{#1}}

\newif\if@cghi
\def\cite{\@cghitrue\@ifnextchar [{\@tempswatrue
	\@citex}{\@tempswafalse\@citex[]}}
\def\citelow{\@cghifalse\@ifnextchar [{\@tempswatrue
	\@citex}{\@tempswafalse\@citex[]}}
\def\@cite#1#2{{$\null^{#1}$\if@tempswa\typeout
	{IJCGA warning: optional citation argument 
	ignored: `#2'} \fi}}

\def\pmb#1{\setbox0=\hbox{#1}
	\kern-.025em\copy0\kern-\wd0
	\kern.05em\copy0\kern-\wd0
	\kern-.025em\raise.0433em\box0}

\def\fnt#1#2{\footnotetext{\kern-.3em
	{$^{\mbox{\scriptsize #1}}$}{#2}}}

\def\fpage#1{\begingroup
\voffset=.3in
\thispagestyle{empty}\begin{table}[b]\centerline{\footnotesize #1}
	\end{table}\endgroup}

\def\runninghead#1#2{\pagestyle{myheadings}
\markboth{{\protect\footnotesize\it{\quad #1}}\hfill}
{\hfill{\protect\footnotesize\it{#2\quad}}}}
\font\tenrm=cmr10
\font\tenit=cmti10 
\font\tenbf=cmbx10
\font\bfit=cmbxti10 at 10pt
\font\ninerm=cmr9

\font\eightrm=cmr8

\def\qed{\hbox{${\vcenter{\vbox{			%HOLLOW SQUARE
   \hrule height 0.4pt\hbox{\vrule width 0.4pt height 6pt
   \kern5pt\vrule width 0.4pt}\hrule height 0.4pt}}}$}}

\def\bsc{{\sc a\kern-6.4pt\sc a\kern-6.4pt\sc a}}	%LATEX LOGO
\def\bflatex{\bf L\kern-.30em\raise.3ex\hbox{\bsc}\kern-.14em 
T\kern-.1667em\lower.7ex\hbox{E}\kern-.125em X} 

\begin{document}

\runninghead{
Spectral correlations in disorderd mesoscopic metals and their relevance
for persistent currents} {
Spectral correlations in disorderd mesoscopic metals and their relevance
for persistent currents}

\normalsize\textlineskip
\thispagestyle{empty}
\setcounter{page}{1}

\copyrightheading{}			%{Vol. 0, No. 0 (1993) 000---000}

\vspace*{0.88truein}

\fpage{1}
\centerline{\bf SPECTRAL CORRELATIONS IN DISORDERED MESOSCOPIC METALS}
\vspace*{0.035truein}
\centerline{\bf AND THEIR RELEVANCE FOR PERSISTENT CURRENTS}
\vspace*{0.37truein}
\centerline{\footnotesize AXEL V\"{O}LKER AND PETER KOPIETZ}
\vspace*{0.015truein}
\centerline{\footnotesize\it 
Institut f\"{u}r Theoretische Physik der Universit\"{a}t G\"{o}ttingen}
\baselineskip=10pt
\centerline{
\footnotesize\it 
Bunsenstr.9, D-37073 G\"{o}ttingen, Germany}
\vspace*{10pt}
\pub{}

\vspace*{0.21truein}
\abstracts{
We use the Lanczos method to calculate 
the variance $\sigma^2 ( E , \phi )$ of the number of energy 
levels in an energy window of width $E$ below the Fermi energy
for non-interacting disordered electrons
on a thin three-dimensional ring threaded 
by an Aharonov-Bohm flux $\phi$.
We confirm numerically that for small $E$
the {\it{flux-dependent}} part of
$\sigma^2 ( E , \phi )$
is well described by the Altshuler-Shklovskii-diagram 
involving two Cooperons. 
However, in the absence of electron-electron 
interactions this result
cannot be extrapolated to energies $E$
where the energy-dependence 
of the average density of states becomes significant. 
We discuss consequences for persistent currents and argue that
for the calculation of the difference between 
the canonical- and grand canonical current
it is crucial to take the electron-electron interaction into account.
}{}{}

\hspace{30mm}

\vspace*{1pt}\textlineskip

The mesoscopic persistent current in a small metal ring
threaded by an Aharonov-Bohm flux $\phi$ has been
predicted long time ago by Hund\cite{Hund38}. 
However, the role of disorder\cite{Buttiker83} and
electron-electron interactions\cite{Ambegaokar90} 
has only been addressed quite recently, and is 
not completely understood\cite{Kopietz93,Vignale94,Altland94}.
The experiment by Levy {\it{et al.}}\cite{Levy90} and the
subsequent experiments by other groups\cite{Chandrasekhar91,Mailly93}
have motivated many recent theoretical works.
There exists now general agreement that
the surprisingly large experimentally observed 
persistent currents\cite{Levy90,Chandrasekhar91} can only be explained by taking
the electron-electron interaction into account. 
Nevertheless, the calculations within models of non-interacting
electrons have lead to new insights into the
nature of mesoscopic disordered systems.
In particular, in a seminal paper Altshuler, Gefen and Imry\cite{Altshuler91}
have shown that for non-interacting electrons in the diffusive
regime the dominant contribution to the average
persistent current arises from striking differences
between the canonical and grand-canonical ensembles.
Because experimentally\cite{Levy90} no external 
leads were attached to the rings,
one should calculate the
{\it{canonical}} persistent current 
 \begin{equation}
   I ( N , \varphi ) =  \frac{- e}{h} 
 \left( \frac{ \partial F ( N , \varphi )}{\partial \varphi} \right)_{N}
 \label{eq:Icandef}
 \; \; \; ,
 \end{equation}
where $F ( N , \varphi )$ is the canonical free energy, 
the flux $\varphi = \phi / \phi_0$ is measured in units of the
flux quantum $\phi_0 = hc /e$, and $-e$ is the charge of the
electron.
Unfortunately, the usual methods of quantum statistical
mechanics are based on the grand canonical formalism, where
the chemical potential $\mu$ is held constant. 
In Ref.\cite{Altshuler91} 
the disorder averaged canonical current
$\overline{I (  N , \varphi ) }$ was therefore related to the variance 
 \begin{equation}
  \Sigma^{2} ( \mu^{\ast} , \varphi )
  = \overline{N^2 ( \mu^{\ast} , \varphi )} -
   \left[ \overline{N ( \mu^{\ast} , \varphi )} \right]^2 
 \label{eq:SigmaNdef}
 \end{equation}
of the particle number $N ( \mu , \varphi )$ in a corresponding
grand-canonical ensemble at a
certain value $\mu^{\ast}$ of the chemical potential.
Here and below the overbar denotes averaging over the disorder.
$\mu^{\ast}$ should be chosen such that 
the disorder- and flux-averaged particle number in the
corresponding grand-canonical ensemble
agrees with the given particle number $N$ in the
original canonical ensemble, i.e.
 \begin{equation}
\int_{0}^{1} d \varphi \overline{ N ( \mu^{\ast}, \varphi )} = N
\label{eq:mustardef}
\; \; \; .
\end{equation}
The expression derived in Ref.\cite{Altshuler91}
can be written as\cite{footnote1}
 \begin{equation}
  \overline{I (  N, \varphi )} -  
  \overline{I_{\rm gc} ( \mu^{\ast} , \varphi) } 
  \approx 
  \frac{- e}{  h }
  \frac{1}{2 
  \kappa ( \mu^{\ast} , \varphi )
  }
  \frac{ \partial }{\partial \varphi} 
  \Sigma^{2} ( \mu^{\ast} , \varphi )
  \; \; \; ,
  \label{eq:AGI}
  \end{equation}
where 
  $\overline{I_{\rm gc} ( \mu^{\ast} , \varphi) } $ is the
average grand-canonical current, and
$ \kappa ( \mu , \varphi ) =  (  
 \partial \overline{N ( \mu , \varphi )} / \partial \mu )_{\varphi}$ 
is the average compressibility.
For non-interacting electrons in the diffusive regime
  $\overline{I_{\rm gc} ( \mu^{\ast} , \varphi) } $ 
is exponentially small; 
the leading interaction contribution to
  $\overline{I_{\rm gc} ( \mu^{\ast} , \varphi) } $
is not negligible and 
has been calculated by Ambegaokar and Eckern\cite{Ambegaokar90}.

Note that Eq.\ref{eq:mustardef} defines $\mu^{\ast}$ as function of $N$, so that
the right-hand side of Eq.\ref{eq:AGI} is indeed a function of $N$ and $\varphi$.
As it stands, Eq.\ref{eq:AGI} is valid even for interacting systems,
and can be used to take the effect of electron-electron interactions 
on the canonical current in a simple but 
non-perturbative way into account\cite{Berkovits96,Voelker96}.
In this work we would like to restrict ourselves to non-interacting
electrons in the diffusive regime. Then $\kappa^{-1} ( \mu^{\ast} , \varphi )$
can be approximated by a flux-independent constant $\Delta$,
the average level spacing at the Fermi energy $\mu^{\ast}$. 
In this way the calculation of the average canonical persistent
current is reduced to the problem of calculating the
{\it{flux-dependent}} part of the variance 
$\Sigma^2 ( \mu, \varphi ) $
of the particle number in a grand canonical ensemble.
To calculate this quantity, let us write
 \begin{equation}
\Sigma^{2} ( \mu , \varphi ) = \lim_{E \rightarrow \infty} \sigma^{2} ( E, \mu , \varphi ) 
 \; \; \; ,
 \end{equation}
with 
 \begin{equation}
\sigma^{2} ( E , \mu , \varphi ) 
 = \int_{\mu -E}^{\mu} d \epsilon
  \int_{\mu -E}^{\mu} d \epsilon^{\prime} 
  K_2 ( \epsilon , \epsilon^{\prime} , \varphi )
 \label{eq:NKdef}
 \; \;\; ,
 \end{equation}
 \begin{equation}
  K_2 ( \epsilon , \epsilon^{\prime} , \varphi )
  = \overline{
 \rho ( \epsilon , \varphi )  
 \rho ( \epsilon^{\prime} , \varphi )  }
 - \overline{ \rho ( \epsilon , \varphi )} \; \;
 \overline{ \rho ( \epsilon^{\prime} , \varphi )} 
 \; \; \; .
 \label{eq:Kdef}
 \end{equation}
Here
 $\rho ( \epsilon , \varphi )  =  \sum_{\alpha} 
 \delta ( \epsilon -  \epsilon_{\alpha} ( \varphi) )$ is
the density of states for a given realization of the disorder.
(The $\epsilon_{\alpha} ( \varphi )$ are the exact solutions of the Schr\"{o}dinger equation
for fixed random potential.)
Note that
$\sigma^{2} ( E , \mu , \varphi ) $ is the 
variance of the number of energy levels in an interval
of width $E$ below the Fermi energy.
Combining Eqs.\ref{eq:AGI}--\ref{eq:Kdef}, we 
obtain for the average canonical persistent current of non-interacting electrons
(ignoring the contribution
$ \overline{I_{\rm gc} ( \mu^{\ast} , \varphi) } $, which is exponentially small
in the diffusive regime)
 \begin{equation}
  \overline{I (  N, \varphi )} \approx  \frac{- e}{  h }
  \frac{\Delta}{2} 
  \frac{\partial}{\partial \varphi}
  \lim_{E \rightarrow \infty}
 \int_{\mu^{\ast} -E}^{\mu^{\ast}} d \epsilon
  \int_{\mu^{\ast} -E}^{\mu^{\ast}} d \epsilon^{\prime} 
  K_2 ( \epsilon , \epsilon^{\prime} , \varphi )
  \; \; \; .
  \label{eq:IKrelation}
  \end{equation}
This expression has been used in Refs.\cite{Altshuler91,Schmid91,Oppen91}
to calculate the average persistent current.
For the covariance function
  $K_2 ( \epsilon , \epsilon^{\prime} , \varphi )$
these authors have substituted
a perturbative approximation
due to Altshuler and Shklovskii\cite{Altshuler86}, 
which for a thin quasi one-dimensional ring (where diffusion is only possible
along the circumference) is given by
 \begin{eqnarray}
  K_2 ( \epsilon , \epsilon^{\prime} , \varphi )
  & \approx & 
  \nonumber
  \\
  &  & \hspace{-14mm}
  \frac{1}{2 \pi^2} {\rm Re} \sum_{ k= - \infty }^{\infty}
  \left\{ 
  \frac{1}{ [4 \pi^2 E_{ c} ( k + 2 \varphi )^2 - i 
  ( \epsilon - \epsilon^{\prime} ) + \Gamma]^2 }
  \right.
  \nonumber
  \\
  & &  
  \left. +
  \frac{1}{ [4 \pi^2 E_{ c} k^2 - i 
  ( \epsilon - \epsilon^{\prime} ) + \Gamma]^2 }
  \right\}
  \label{eq:ASexpression}
  \; \; \; .
  \end{eqnarray}
The first term in the curly braces is due to
the two-Cooperon diagram, while the second term
is due to the two-Diffuson diagram\cite{Altshuler86}.
Here $E_c$ is the Thouless energy\cite{Altshuler91}, and $\Gamma$ 
is a cutoff energy 
that has been introduced  by hand into the non-interacting model. 
For non-interacting electrons 
non-perturbative effects due to higher-order terms 
give rise to contributions to $\Gamma$ of the order of
the average level spacing\cite{Efetov83,Dupuis91}.
In the presence of electron-electron interactions, $\Gamma$
should take inelastic processes
approximately into account\cite{Schmid91}. 
Eq.\ref{eq:ASexpression} is believed to be the dominant
contribution to $K_2$ in the regime
$ \Delta 
\raisebox{-0.5ex}{$\; \stackrel{<}{\sim} \;$}
| \epsilon - \epsilon^{\prime} | 
\raisebox{-0.5ex}{$\; \stackrel{<}{\sim} \;$}
\hbar / \tau$, where $\tau$ is the elastic lifetime.

It is tempting
to substitute Eq.\ref{eq:ASexpression} into Eq.\ref{eq:IKrelation}
and obtain in this way a simple analytic result for the average canonical
persistent current\cite{Altshuler91,Schmid91,Oppen91}.
However, such a procedure is based on a hidden assumption,
which apparently has not been noticed in the recent literature on
persistent currents: 
In the  derivation of Eq.\ref{eq:ASexpression}
it is implicitly assumed that 
in the energy window of interest the
{\it{energy-dependence of the average density of states can be neglected}}.
On the other hand, 
according to Eq.\ref{eq:IKrelation}
the limit $E \rightarrow \infty$ has to be taken
in order to obtain the physical persistent current,
so that we need to know the function
$K_2 ( \epsilon , \epsilon^{\prime} , \varphi )$
for all energies 
$\epsilon$ and $\epsilon^{\prime}$ below the Fermi energy.
Obviously, in this regime
the energy-dependence of the
average density of states
$\overline{\rho ( \epsilon , \varphi )}$
cannot be ignored.
Of course, if we proceed by performing the above substitution anyway,
it is easy to show that
(see Eq.\ref{eq:P0mlim} below)
\begin{equation}
\frac{\partial}{\partial \varphi}
\Sigma^2 (
\mu , \varphi ) =
\frac{\partial}{\partial \varphi}
\sigma^2 ( E , \mu , \varphi ) 
\; \; \; , \; \; \; 
\mbox{for all $ E  
\raisebox{-0.5ex}{$\; \stackrel{>}{\sim} \;$}
E_c $}
\; \; \; .
\label{eq:saturate}
\end{equation}
Thus, at the first sight it seems that 
the limit $E \rightarrow \infty$
in Eq.\ref{eq:IKrelation} is trivial
and the above
substitution is justified.
However, this way of reasoning is clearly not {\it{self-consistent}}, because
the result \ref{eq:saturate} depends 
via Eq.\ref{eq:ASexpression}
on the {\it{assumption of a constant density of states}}.
In this work we shall carefully examine this point numerically.
Our main result is that for non-interacting electrons
Eq.\ref{eq:saturate} is {\it{not}} correct, 
so that
the physical persistent current cannot be calculated by
simply substituting the Altshuler-Shklovskii result \ref{eq:ASexpression}
into Eq.\ref{eq:IKrelation}.
We also present a numerical test of 
Eq.\ref{eq:ASexpression}. 
In a three-dimensional system Altshuler-Shklovskii scaling has recently
been reproduced numerically by Braun and Montambaux\cite{Braun95}, but
these authors examined only the flux-independent part of
$\sigma^2 ( E , \mu , \varphi )$.

Using the Lanczos method, we have 
numerically calculated the exact energy levels $\epsilon_{\alpha} ( \varphi )$
of the spinless nearest-neighbor tight-binding Anderson 
Hamiltonian\cite{Anderson58} with
diagonal disorder. 
The site-diagonal random potentials are
assumed to be box-distributed in the interval $[ -w /2 , w/2]$.
The Aharonov-Bohm flux is taken into account
by choosing flux-dependent hopping energies\cite{Bouchiat89}
in the $\pm x$-directions,
$t_{\pm x} = t e^{\pm 2 \pi i \varphi / N_x }$, where $N_x$ is the
number of lattice sites in the direction $x$.
For convenience, all energies 
will be measured in units of the
hopping energy $t$, i.e. we formally set $t = 1$. 
For a quasi one-dimensional ring with
$N_x = 20$, $N_y = N_z =5$ we found diffusive 
behavior for $w = 2.5$. 
However, in order to reduce the statistical errors 
so that even the flux-dependent part
of $\sigma^2 ( E , \mu , \varphi )$
can be resolved, it is necessary to 
average over an extremely large number 
(typically $20000$)
of statistically independent realizations of the random potential.
Our calculations were performed with the help of a 
parallel code\cite{Holm95} 
on a cluster of up to $30$ work-stations.

\begin{figure}[htpb]
\vspace*{13pt}
\hspace{16mm}
\epsfysize7cm 
\epsfbox{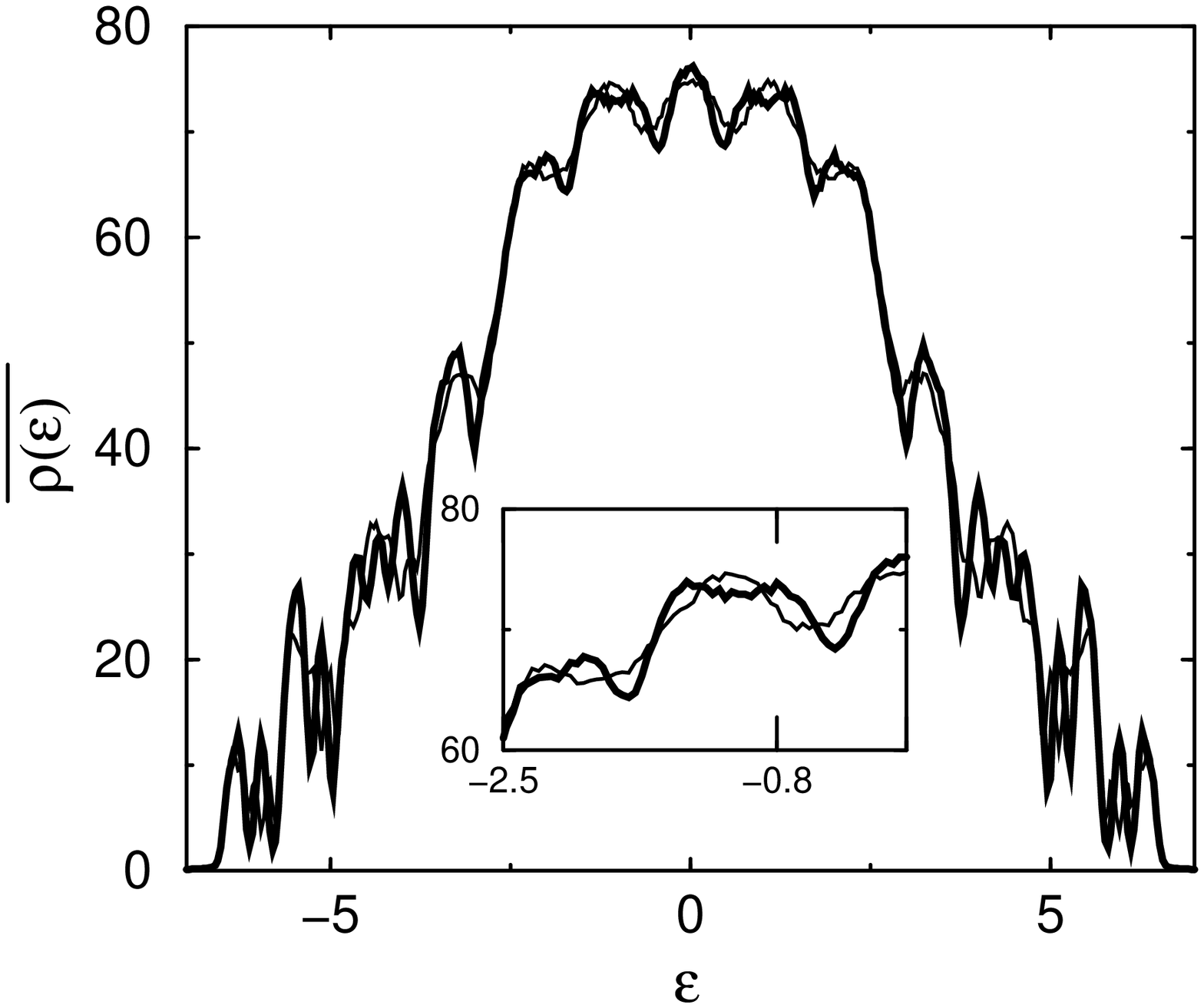}
%\vspace{1cm}
\fcaption{
Average density of states for a $20 \times 5 \times 5$ lattice.
We used $w = 2.5$ and averaged over $20000$ realizations of the disorder.
Thick line: $\varphi = 0$; thin line: $\varphi = 1/4$.
The inset shows part of the central regime on a larger scale, 
together with our choice of the chemical potential $\mu = -0.8$.} 
\label{fig:dos}
%\centerline{\vbox{\hrule width 5cm height0.001pt}}
%\vspace*{1.4truein}		%ORIGINAL SIZE=1.6TRUEIN x 100% - 0.2TRUEIN
%\centerline{\vbox{\hrule width 5cm height0.001pt}}
%\vspace*{13pt}
%\fcaption{Labeled tree {\footnotesize\it T}.}
\end{figure}

Because the Altshuler-Shklovskii result \ref{eq:ASexpression}
is based on the assumption that in the energy window of interest
the average density of states $\overline{\rho (\epsilon , \varphi )}$
can be approximated by a constant,
it is important to choose the chemical potential $\mu$
properly.  In Fig.\ref{fig:dos} we show the numerical result
for the average density of states for a typical set of parameters.
Obviously  $\mu = - 0.8$ is a good choice, because
then the Fermi energy lies on the right-hand side of a broad maximum, so that
the width of the interval
$[ \mu - E , \mu]$ where $\overline{\rho ( \epsilon , \varphi )}$ 
can be approximated  by a constant is as large as possible.

\begin{figure}[htbp]
\vspace*{13pt}
\hspace{16mm}
\epsfysize7cm 
\epsfbox{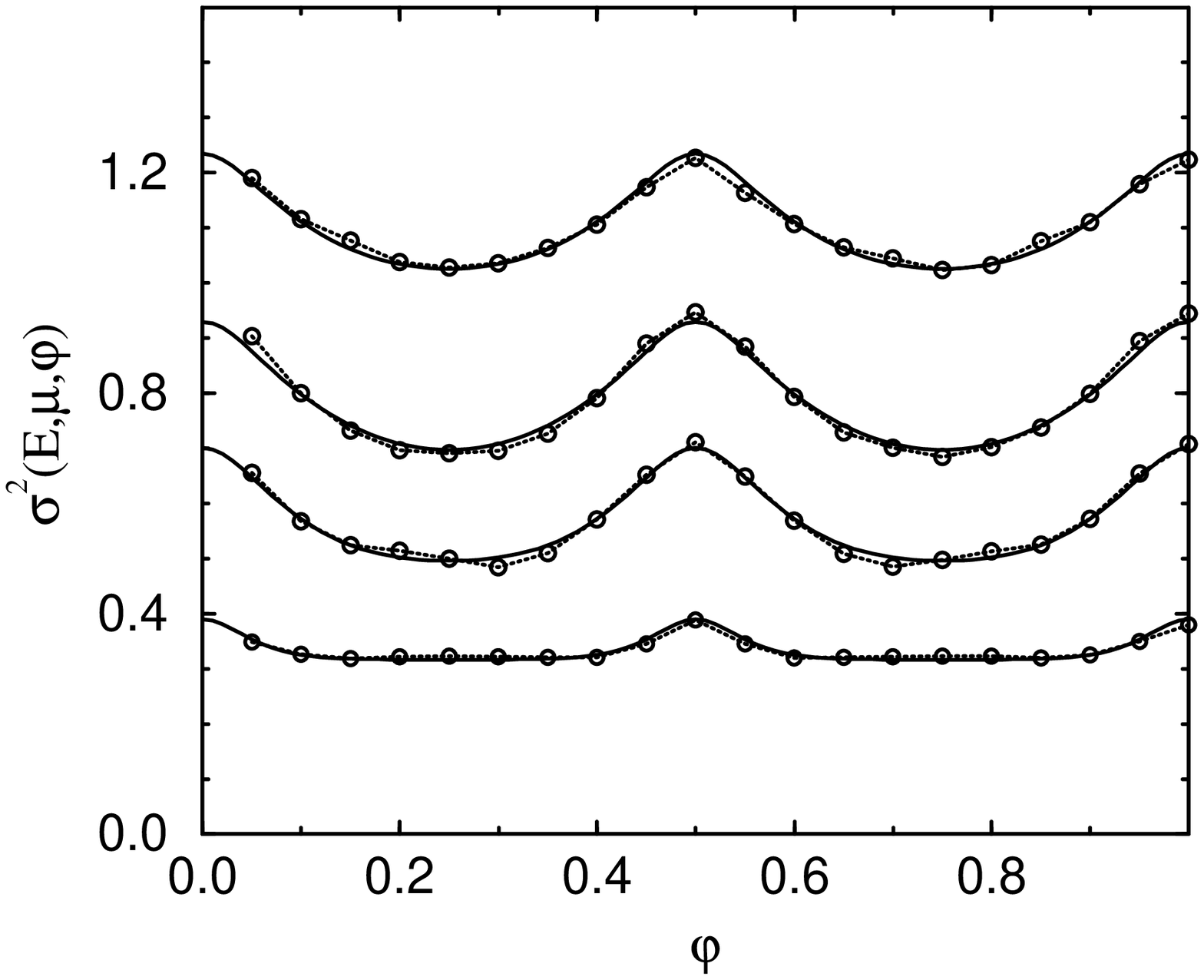}
%\vspace{1cm}
\fcaption{
The variance $\sigma^2 ( E , \mu , \varphi )$ as function of
$\varphi$ for different values of $\tilde{E} = E / E_c$ for $\mu = -0.8$
and $E_c = 0.01$. From bottom to top:
$\tilde{E} = 1$; $\tilde{E} = 5$; 
$\tilde{E} = 15$; $\tilde{E} = 60$.
The solid lines 
are theoretical curves according to Eq.\ref{eq:P0mE} with
$\tilde{\Gamma} = 0.55$.
We have chosen the flux-average 
$\sigma^{2}_0  ( \tilde{E} )$ as a fitting
parameter. See Fig.\ref{fig:zerothF}
for a numerical calculation of
$\sigma^{2}_0  ( \tilde{E} )$.
}
\label{fig:AS2}
%\centerline{\vbox{\hrule width 5cm height0.001pt}}
%\vspace*{1.4truein}		%ORIGINAL SIZE=1.6TRUEIN x 100% - 0.2TRUEIN
%\centerline{\vbox{\hrule width 5cm height0.001pt}}
%\vspace*{13pt}
%\fcaption{Labeled tree {\footnotesize\it T}.}
\end{figure}

\begin{figure}[htbp]
\vspace*{13pt}
\hspace{16mm}
\epsfysize7cm 
\epsfbox{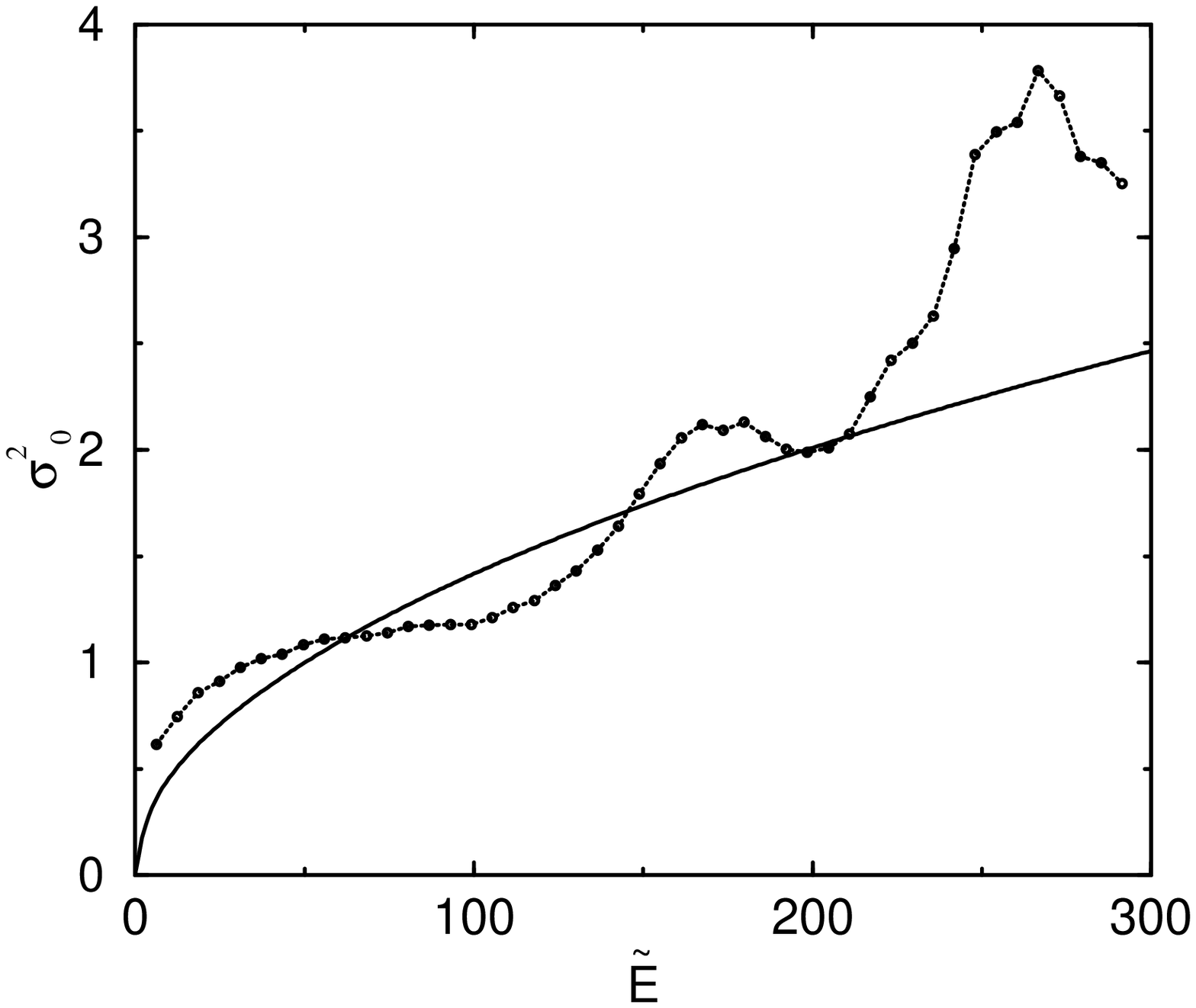}
%\vspace{1cm}
\fcaption{
Flux average $\sigma^2_0$ of
$\sigma^2( E , \mu , \varphi )$ as function of
$\tilde{E} = E / E_c $.
The parameters are chosen as in Figs.\ref{fig:dos} and \ref{fig:AS2}.
The solid line is the theoretical prediction \ref{eq:P00E}
with $\tilde{\Gamma} = 0.55$.
}
\label{fig:zerothF}
%\centerline{\vbox{\hrule width 5cm height0.001pt}}
%\vspace*{1.4truein}		%ORIGINAL SIZE=1.6TRUEIN x 100% - 0.2TRUEIN
%\centerline{\vbox{\hrule width 5cm height0.001pt}}
%\vspace*{13pt}
%\fcaption{Labeled tree {\footnotesize\it T}.}
\end{figure}

Next, let us test the Altshuler-Shklovskii prediction
for the variance 
$\sigma^2( E , \mu , \varphi )$ of the
number of energy levels in an energy window of width $E$ below the
Fermi energy. For better comparison with the numerical
results, it is convenient to expand
the theoretical prediction 
for $\sigma^2 ( E , \mu , \varphi )$ 
into a Fourier series.
Substituting Eq.\ref{eq:ASexpression} into
Eq.\ref{eq:NKdef}, we obtain
 \begin{equation}
\sigma^2 ( E , \mu , \varphi ) 
  =  \sigma^2_0 ( \tilde{E} ) 
  +  \sum_{m=1}^{\infty} 
 \sigma^2_{2m} ( \tilde{E} ) \cos ( 4 \pi m \varphi )
 \label{eq:PE0fourier}
 \; \; \; ,
 \end{equation}
where for $m \geq 1$
 \begin{eqnarray}
 \sigma^2_{2m} ( \tilde{E} )  & = & \frac{2}{\pi^2 m}
 \left\{  
 \exp \left[ -m \tilde{\Gamma}^{1/2} \right]
 - 
  \exp \left[
 -  \frac{m}{\sqrt{2}}
 \left( 
  \sqrt{ \tilde{E}^2 + \tilde{\Gamma}^2 } + 
  \tilde{\Gamma}
  \right)^{1/2} 
  \right]
 \nonumber
 \right.
 \\
 & & 
 \left.
 \hspace{3cm}
 \times \cos
 \left[ \frac{m}{\sqrt{2}}
 \left( 
   \sqrt{\tilde{E}^2 + \tilde{\Gamma}^2 } - \tilde{\Gamma} \right)^{1/2}  
  \right]
 \right\}
 \; \; \; ,
 \label{eq:P0mE}
 \end{eqnarray}
and the flux average is 
 \begin{eqnarray}
 \sigma^2_0 ( \tilde{E} ) & = & \frac{2}{ \pi^2}
 \left[ 
 \frac{1}{\sqrt{2}}
 \left( 
  \sqrt{ \tilde{E}^2 + \tilde{\Gamma} } + \tilde{\Gamma} \right)^{1/2}  
    -  {\tilde{\Gamma}}^{1/2} \right]
 \nonumber
 \\
 & + & \sum_{m=1}^{\infty} \sigma^2_{2m} ( \tilde{E})
 \label{eq:P00E}
 \; \; \; .
 \end{eqnarray}
Here $\tilde{E} = E / E_c$, and $\tilde{\Gamma} = \Gamma / E_c$.
Note that $E_c$ (and hence $\tilde{E}$) implicitly depends on $\mu$.
For $\tilde{E} \gg 1 \gg \tilde{\Gamma}$
Eqs.\ref{eq:P0mE} and \ref{eq:P00E} reduce to
 \begin{eqnarray}
 \sigma^2_{2m} ( \tilde{E}  ) & \sim   &
 \frac{ 2 }{\pi^2 m} 
 \exp \left[ - m \tilde{\Gamma}^{1/2}  \right] 
 \; , \;  m =1,2, \ldots
 \label{eq:P0mlim}
 \\
 \sigma^2_0 ( \tilde{E} ) & \sim   &
 \frac{  \sqrt{2}}{ \pi^2} \tilde{E}^{1/2}
 + \frac{2}{\pi^2 } \ln [ \tilde{\Gamma}^{-1/2} ]
 \label{eq:P00lim}
 \; \; \; .
 \end{eqnarray}
Hence, the higher harmonics $\sigma^2_{2m} ( \tilde{E}  ) $
become independent of $\tilde{E}$ as soon as 
$\tilde{E } 
\raisebox{-0.5ex}{$\; \stackrel{>}{\sim} \;$} 1$,
implying the validity of
Eq.\ref{eq:saturate}.
Physically this means that the dominant 
contribution to the flux-dependent part of 
$\sigma^2 ( E ,  \mu , \varphi )$
is due to states with energies in an interval of width $E_c$ below 
the Fermi level. In Fig.\ref{fig:AS2} we show our numerical
results for  the flux-dependent part of
$\sigma^2 ( E ,  \mu , \varphi )$ 
for different values of $\tilde{E} = E / E_c$.
The flux-oscillations  with a fundamental
period of $\phi_0 / 2$ are  clearly visible, and the amplitude is in perfect agreement
with the diagrammatic calculation based on the
two-Cooperon diagram of Altshuler and Shklovskii\cite{Altshuler86}.
Note also that the amplitude of the flux modulation does not
change when the energy $E$ is increased beyond the Thouless energy.
To the best of our knowledge, this is the first numerical
confirmation  that the flux-dependence 
of $\sigma^2 ( E ,  \mu , \varphi )$
is indeed correctly described by the two-Cooperon diagram
given in Ref.\cite{Altshuler86}.
Our numerical data for the
{\it{flux-independent}} part
$\sigma^2 ( E ,  \mu , \varphi )$ is shown
in Fig.\ref{fig:zerothF}.
Although the overall quantitative agreement is not as good as
in the case of the flux-dependent part, the
quasi-one-dimensional $\sqrt{E}$-behavior 
is clearly visible.
It should be kept in mind, however, that the Altshuler-Shklovskii 
calculation
is only valid for energies
$E \raisebox{-0.5ex}{$\; \stackrel{<}{\sim} \;$} \hbar / \tau $.
Using simple second order perturbation theory, we estimate
$\hbar / \tau \approx 0.5$ in the parameter regime relevant
to the figures.  With $E_c \approx 0.01$, the theoretical
curve in Fig.\ref{fig:zerothF} should be 
quantitatively accurate for $\tilde{E}
\raisebox{-0.5ex}{$\; \stackrel{<}{\sim} \;$} 50 $.

\begin{figure}[htbp]
\vspace*{13pt}
\hspace{16mm}
\epsfysize7cm 
\epsfbox{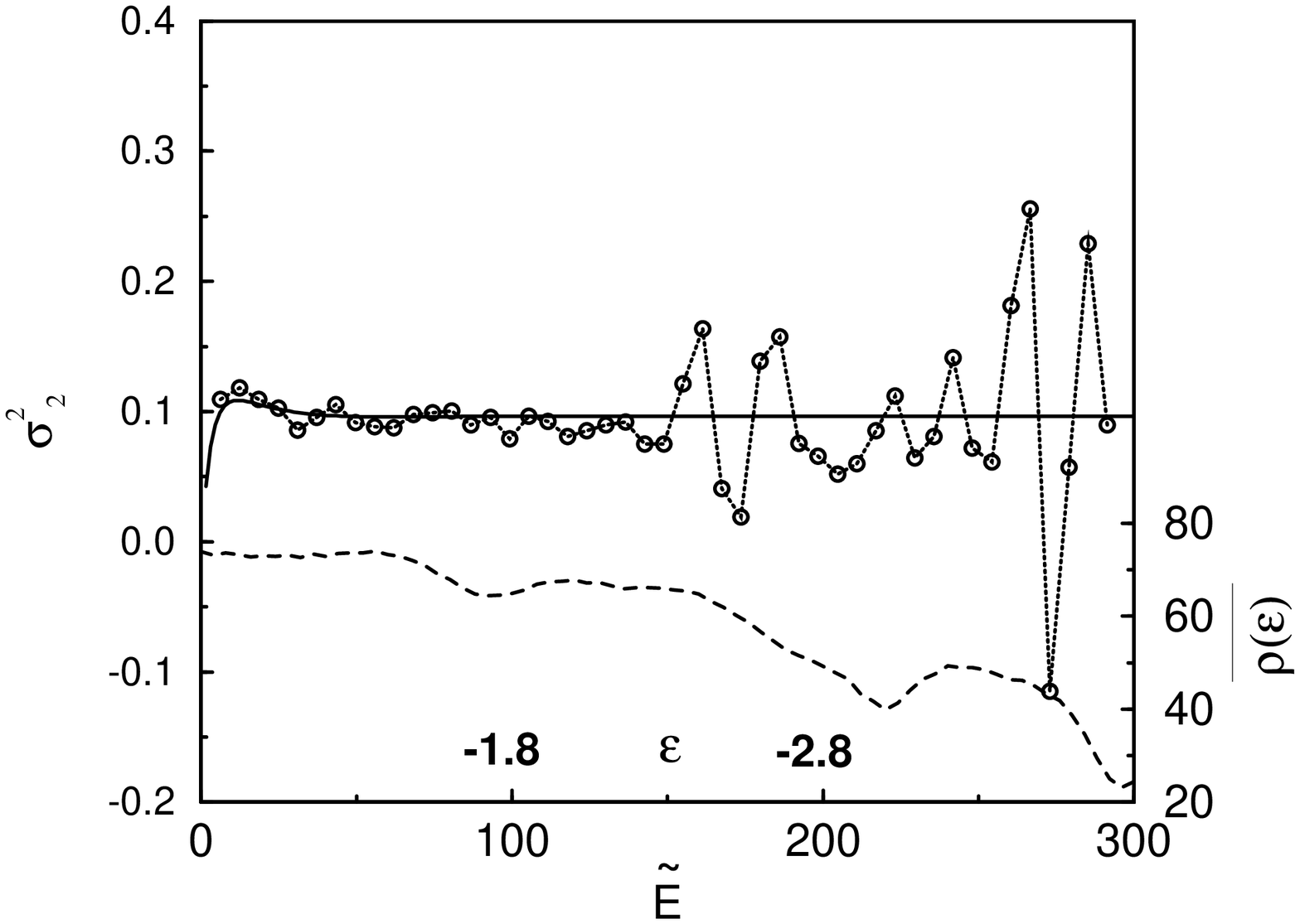}
%\vspace{1cm}
\fcaption{
Second Fourier component $\sigma^2_2$ of
$\sigma^2 ( E , \mu ,  \varphi )$ as function of
$\tilde{E}$.
The parameters are chosen as in Figs.\ref{fig:dos} and \ref{fig:AS2}.
The solid line is 
the theoretical prediction \ref{eq:P0mE} with $\tilde{\Gamma} = 0.55$.
The upper labels on the horizontal axis and the right
vertical scale refer to the
corresponding average density of states 
$\overline{\rho ( \epsilon )}$
at $\varphi =0$ (dashed line).
}
\label{fig:sat}
%\centerline{\vbox{\hrule width 5cm height0.001pt}}
%\vspace*{1.4truein}		%ORIGINAL SIZE=1.6TRUEIN x 100% - 0.2TRUEIN
%\centerline{\vbox{\hrule width 5cm height0.001pt}}
%\vspace*{13pt}
%\fcaption{Labeled tree {\footnotesize\it T}.}
\end{figure}

The theoretical results \ref{eq:P0mE}--\ref{eq:P00lim}
have been derived under the assumption that in the
energy interval of interest the  average density of states can be
approximated by a constant.
The important question is now whether
it is allowed to use the small-$E$ regime 
as a basis for the extrapolation
of the {\it{flux-dependent}} part of $\sigma^{2} ( E , \mu , \varphi )$
to $E \rightarrow \infty$. According to Eq.\ref{eq:IKrelation}
this extrapolation is necessary 
to obtain the persistent current.
Note that in Refs.\cite{Altshuler91,Schmid91,Oppen91} such a procedure is adopted
without further comment.
In Fig.\ref{fig:sat}
we have plotted our numerical result for the
second harmonic $\sigma^2_2  ( \tilde{E} )$ as function of energy.
Although for sufficiently small $\tilde{E}$ we obtain
excellent agreement with
the theoretical result \ref{eq:P0mE},
at $\tilde{E} \approx 150$
the numerical data begin to deviate strongly
from the theoretical curve.
Note that this energy 
cannot be identified with
$\hbar / ( \tau E_c ) \approx 50$ --
instead, it is the energy scale
where the average density of states
(shown as dashed line in Fig.\ref{fig:sat})
starts to deviate significantly from its value
at the Fermi energy.
In contrast, according to the theoretical
prediction \ref{eq:P0mlim}
the non-zero harmonics $\sigma^2_{2m} ( \tilde{E} )$
should saturate for $\tilde{E} 
\raisebox{-0.5ex}{$\; \stackrel{>}{\sim} \;$} 1 $, 
and nothing special should happen at $\tilde{E} = \hbar / ( \tau E_c ) \approx 50$.
Indeed, at $ \tilde{E}  \approx 50$ 
our numerical result for $\sigma^2_2 ( \tilde{E} )$ still agrees
well with the theoretical prediction.
However, as soon as the average density of states deviates
from its value at the Fermi energy,
our numerical data in Fig.\ref{fig:sat}
clearly disagree with the theoretical curve. 
We therefore conclude that the extrapolation of  
Eq.\ref{eq:P0mE}
to $E \rightarrow \infty$ is not possible.
Hence, at least for non-interacting electrons, 
{\it{the average persistent
current cannot be calculated by simply substituting 
the Altshuler-Shklovskii result 
for $\sigma^2 ( E , \mu , \varphi )$
into the formula \ref{eq:IKrelation}
and taking the limit $E \rightarrow \infty$ to
obtain the variance
$\Sigma^2 ( \mu , \varphi )$
of the total particle number.}}
The physical reason is that 
$\Sigma^2 ( \mu , \varphi )$
depends  on all energies $\epsilon_{\alpha} ( \varphi ) \leq \mu$. 
Because in the diffusive regime energy levels
tend to repel each other\cite{Efetov83}, 
the fluctuations of energy levels
$\epsilon_{\alpha} ( \varphi )$ deep inside the
Fermi sea induce also changes in the energy levels in the
vicinity of the Fermi energy. 
Thus, the variance of the total particle number
depends on the statistical properties of the entire spectrum below the
 Fermi energy, so that
the energy-dependence of the density of states cannot be ignored.

Realistic models of non-interacting electrons
certainly should have  energy-dependent densities of states, so that
for non-interacting electrons
the flux-dependent part of the variance $\Sigma^2 ( \mu , \varphi )$
of the total particle number cannot be
calculated  from the Altshuler-Shklovskii diagram.
In other words, in the absence of electron-electron interactions
the second term in Eq.\ref{eq:AGI} {\it{is not determined by the
statistical properties of the spectrum in the vicinity
of the Fermi energy}}.
On the other hand, in the presence
of electron-electron interactions
we expect a completely different scenario:
in this case Eq.\ref{eq:saturate}
should be {\it{at least qualitatively correct}},  
because in an interacting many-body system 
weakly damped propagating quasi-particles
exist only in the vicinity
of the Fermi energy.
But only quasi-particles 
that can propagate coherently around  the ring
can probe the sensitivity to twists in the
boundary conditions (and thus contribute to the flux-dependence
of the particle number variance), 
so that we expect that electron-electron interactions will
eliminate the contribution from
many-body states with energies deep inside the
Fermi sea.
In other words the difference
$\overline{I ( N , \varphi )} - \overline{I_{\rm gc} ( \mu^{\ast} , \varphi )}$ in
Eq.\ref{eq:AGI} should be extremenly sensitive to electron-electron interactions
in the sense that only if the damping of the quasi-particles is taken into account,
the flux-dependent part of $\Sigma^{2} ( \mu^{\ast} , \varphi )$ is determined by
the spectrum close to the Fermi energy (see Eq.\ref{eq:saturate}).
In this sense the procedure adopted
in Refs.\cite{Altshuler91,Schmid91,Oppen91} was physically correct\cite{footnote2},
although within a model of non-interacting electrons
it cannot be justified.
Note, however, in the presence of electron-electron interactions
the contribution $\overline{I_{\rm gc} ( \mu^{\ast} , \varphi )}$ in
Eq.\ref{eq:AGI} is not negligible, and it is not clear whether the
persistent current is still dominated by the difference between the
canonical and the grand-canonical current.

We would like to emphasize that at this point the above scenario
should be considered as a plausible hypothesis, 
which implicitly\cite{footnote2} has also been made
in Refs.\cite{Altshuler91,Schmid91,Oppen91}.
We suspect that in an interacting many-body system the microscopic mechanism
which eliminates contributions 
from states deep inside the Fermi sea
is closely related to the 
phenomenological cutoff parameter $\Gamma$
in Eq.\ref{eq:ASexpression}.
A perturbative calculation of $\Gamma$
in an interacting mesoscopic conductor 
has recently been reported  by Blanter\cite{Blanter96},
but in the present context more accurate calculations
are needed, 
which take also the energy-dependence of $\Gamma$ into account.
Although we have started from a model of non-interacting electrons,
we have arrived at the problem of electron-electron interactions.
We believe that a satisfactory and generally accepted solution
of the long-standing persistent current problem
will only be obtained if the role of electron-electron interactions
in mesoscopic disordered conductors is more thoroughly understood.

To conclude, let us briefly summarize our main results:
First of all, we have numerically confirmed that
for a thin three-dimensional ring in the diffusive regime
the variance $\sigma^2 ( E , \mu , \varphi )$ 
of the number of energy levels in a small interval of width $E$
can indeed be described by the perturbative expression derived
by Altshuler and Shklovskii\cite{Altshuler86}.
Our second important result is the observation that
for non-interacting electrons
the flux-derivative  $ \partial \Sigma^{2} ( \mu , \varphi )/ \partial \varphi$ of the
variance of the {\it{total}} number of energy levels below the Fermi energy
cannot be obtained from 
the Altshuler-Shklovskii expression for
$\sigma^2 ( E , \mu , \varphi )$. We have argued that
electron-electron interaction should modify this result, but
a microscopic proof of this conjecture
requires a better understanding
of dephasing in disordered mesoscopic conductors.
We hope to come back to this point in a future publication.

\nonumsection{Acknowledgements}
%\vspace{0.2cm}
\noindent
This work was financially supported by the Deutsche Forschungsgemeinschaft
(SFB 345).
We would like to thank J. A. Holm for providing us with his
``Master-Slave Implementation
for Parallel Virtual Machines'', which 
was crucial to run our numerical simulation efficiently
on a work-station cluster.

\nonumsection{References}

\end{document}